\documentclass[11pt,twoside]{article}
\usepackage{./schechter}

\aspSuppressVolSlug
\resetcounters

\begin{document}

\title{The Contribution of Faint, Failed and Defunct Stars to
the "Stellar" Masses of Galaxies}
\author{Paul L. Schechter$^1$ 
\affil{$^1$MIT Kavli Institute, Cambridge, Massachusetts, USA; \email{schech@mit.edu}}}

\paperauthor{Paul L. Schechter}{schech@mit.edu}{ORCID_Or_Blank}{MIT}{Kavli Institute}{Cambridge}{MA}{02138}{USA}

\begin{abstract}
A substantial fraction the stellar mass attributed to galaxies is
invisible: stars close to the hydrogen burning limit, brown dwarfs,
white dwarfs, neutron stars and black holes.  These constituents do,
however, gravitationally micro-lens background quasars, thereby
permitting measurement of the total stellar contribution to the mass
surface density along the line of sight.  We report the results of
such a measurement using a sample of ten quadruply lensed quasars.
We discuss the prospects for improving upon this measurement with
a larger sample and describe efforts to find new quadruple lenses.
If we invert our argument and take the stellar mass to be known,
we derive a value for the fraction of the dark halo in MaCHOs (including
$\sim20 M_\odot$ primordial black holes) of somthing less than 10\%, confirming
the widely ignored result of Mediavilla et al (2009).
\end{abstract}

\section{Introduction}

Wikipedia defines a galaxy as ``a gravitationally bound system of
stars, stellar remnants, interstellar gas, dust, and dark matter.''
Establishing the relative proportions of each is one of the great
challenges of extragalactic astronomy.  Hundreds of papers
have been written on the stellar masses of galaxies, but
the accuracy with which they can be measured is very much a
matter of debate.
\medskip

{\narrower\noindent
\emph{Nobody ever measures the stellar mass.  That is not a measurable
thing, it's an inferred quantity.  You measure light, OK?  You can
measure light in many bands but you infer stellar mass.  Everybody
seems to agree on certain assumptions that are completely unproven.}
-- Carlos Frenk, 2017 May 15\footnote {\url{http://online.kitp.edu/galhalo-c17/panel1/rm/jwvideo.html} (44:48)} 
\par}

\medskip
The difficulties in measuring stellar masses are magisterially 
examined in a recent paper by Newman et al (2017) comparing stellar
masses for three galaxies computed separately using stellar dynamics,
gravitational macro-lensing and stellar population synthesis.
Variants of each of the three approaches are considered.  The
principal source of uncertainty appears to be the contribution of dark
matter in the first two methods and the low mass cutoff in the stellar
initial mass function in the third.  In a paper on a different galaxy by
three of the same authors, Conroy et al (2017) say:

\medskip
{\narrower\noindent
\emph{To illustrate the sensitivity of the total mass to the cutoff, for a
single power law with $\alpha = 2.7$, the mass-to-light ratio is 70\%
higher if the cutoff is $0.05 M_\odot$ compared to $0.08 M_\odot$.}
\par}

\medskip
Frenk is wrong to say that \emph{everyone} agrees to certain
assumptions.  Schechter and Wambsganss (2004) describe a micro-lensing
method for measuring the stellar-to-dark surface mass density in
galaxies that macro-lens quasars.  The technique measures the \emph
{graininess} of the gravitational potential, to which faint stars,
brown dwarfs and stellar remnants all contribute, invisible though
they might be.  That approach has been refined (Schechter et al 2014)
yielding a stellar surface mass density for a sample of ten lensing
galaxies that is a factor of $1.23 \times e^{\pm 0.47}$ greater than
that of a Salpeter IMF with a $0.10 M_\odot$ cutoff.  The uncertainty
is dominated by the small sample size.

In what follows we review the micro-lensing technique for measuring
stellar masses and then report on efforts to increase the size of
the lensing galaxy sample.

\section{Stellar Masses from Micro-lensing}

\subsection{Flux Ratio Anomalies}

Witt et al (1995) argued that gravitational micro-lensing is a
``universal'' phenomenon in gravitationally lensed quasars and that it
was responsible for the flux ratio anomaly observed in MG0414+0534,
where one of the images was (and still is today) more than a magnitude
fainter than expected from the macro-model for the gravitational
potential.

\vskip2.6truein
\includegraphics{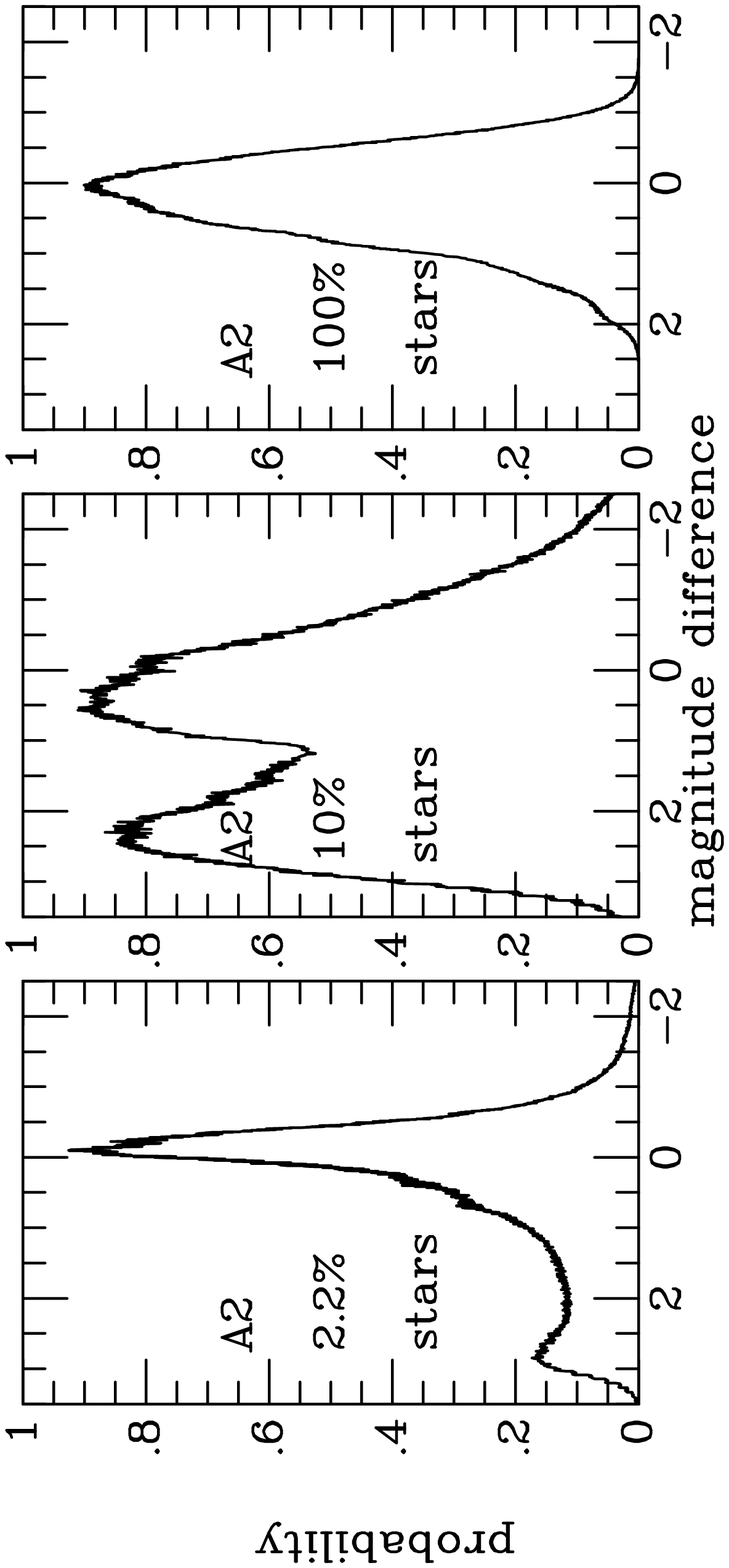}
{\narrower\baselineskip=11pt\noindent
Figure 1.  Probability distribution for the ratio
of observed to macro-model flux (expressed as a magnitude difference)
at three different stellar mass fractions  for 
the $A2$  image of the quadruple lens PG 1115+080.
The different shapes of the distributions
permit determination of  the stellar mass fraction.
\par}
\medskip
While one might expect the amplitudes of those flux ratio anomalies to
increase with increasing stellar surface mass densities, Schechter and
Wambsganss (2002) showed the dependence is not monotonic.  This is
illustrated in figure 1, where the micro-lensing probability density
distributions are shown for three different stellar mass fractions for
the A2 image of PG1115+080, the first quadruply lensed quasar.
The micro-lensing is less strong when all of the surface density
is in stars than when only 10\% is in stars.

One might also expect the flux ratio anomalies to depend upon
the masses of the stars involved, but they have been shown to be
extremely insensitive to the distribution of stellar masses (Schechter
et al 2004), subject to the condition that the emission comes from
regions small compared to the Einstein radii of the micro-lensing stars.
To excellent approximation they depend only on the surface mass density.

To understand the micro-lensing fluctuations, one must remember that
images appear wherever the light travel time from the quasar to the
observer has a stationary point. In the
absence of a lens, there will only be one image, a minimum of the
light travel time.  A galaxy with a sufficiently elliptical potential
produces two minima, two saddlepoints and a maximum, the last of which
is almost always infinitely demagnified.  The stars that lie close to
each macro-image produce micro-minima and micro-saddlepoints, breaking the
macro-images into micro-images (Paczynski 1986).

\subsection{Twinkling Quasars}

The micro-images are the gravitational analog of the speckles produced
by the Earth's atmosphere.  The movement of the stars within the
galaxy and of the galaxy relative to the quasar causes the speckle
pattern to change, and with that the brightness of the speckles.
The quasars scintillate, just as stars do.

This suggests a straighforward approach to measuring the surface mass
density of the A2 image in PG1115+080.  Carry out repeated photometric
observations, accumulate a histogram of fluxes, and compare it
with the panels in figure 1.

Unfortunately the timescale for gravitational scintillation in lensed
quasars is of order ten years.  As is often done in astronomy, one can
substitute single epoch observations of a large number of similar
objects for many observations of one object.

\subsection{Estimating Stellar Surface Mass Density}

One proceeds from stellar fluxes to stellar masses as follows:
\begin{enumerate}
\item Measure fluxes from images.

\item Measure/estimate stellar surface brightness at position
of each quasar image.

\item Make an initial guess of  $M/L,$  the stellar mass-to-light ratio.

\item Carry out micro-lensing simulations and compute
micro-lensing probability distributions based on adopted $M/L$.

\item Assign a figure-of-merit to measure
consistency of fluxes and probability distributions.

\item Make a new guess of $M/L$ and iterate.
\end{enumerate}

One can sidestep the problems associated with measuring surface
brightness at the position of the quasar images and bringing
the surface brightnesses to a common bandpass and epoch by using the
stellar mass fundamental plane (Hyde and Bernardi 2009).  Instead
of $M/L$, the free parameter is a factor ${\cal F}$ by which one
multiplies an adopted stellar mass fundamental plane to obtain the best
agreement with the observed fluxes.
\section{Wanted: More Quadruply Lensed Quasars}

The multiplicative uncertainty in the Schechter et al (2014) result, a
factor of 1.6, would be reduced to a factor of roughly 1.3 if the
sample size were quadrupled from ten to forty.  Given how long it took
to assemble the first ten, one might be tempted to skip the remainder
of this contribution.  But the rate at which new quadruple lenses
are being discovered has accelerated over the past 3 years with the
availability of the VST-ATLAS (Shanks et al 2015), DES (Abbott et al 2016), and
PanSTARRS (Chambers et al 2016) surveys.

In figure 2 we show images of eight of fifteen quadruply
lensed quasars known to the author to have been discovered in the past
three years.  Two of the images are from VST-ATLAS, three are from DES
and three are from PanSTARRS.  The teams discovering these systems
included Lin et al (2016), Agnello et al (2017), Berghea et al (2017),
Ostrovski et al (2017), Lucey et al (2017), Anguita et al (private
communication) and Schechter et al (to be published).

\vskip1.8truein
\includegraphics{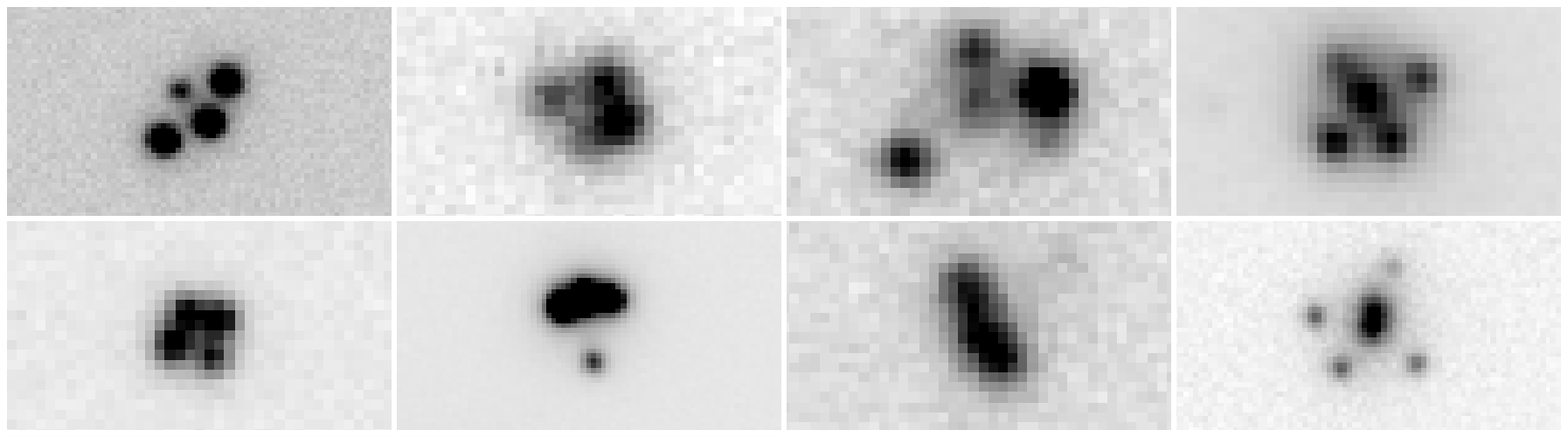}
{\narrower\baselineskip=11pt\noindent
  Figure 2.  Eight quadruply lensed quasars discovered in the past
  3 years.  Images for the first two are taken from the VST-ATLAS
  survey, the next three from the DES, and the last three from
  PanSTARRS.  The scales for the surveys are $0\farcs21$, $0\farcs26$
  and $0\farcs25$ per pixel, respectively.  The second image is in
  Sloan $r$, with all the rest in Sloan $i$.
  \par}
\medskip  

There is reason to think that the acceleration of the discovery rate
will continue.  Until now lensed quasars have been found by first
looking for quasar-colored objects and then resolving them into
multiple images.  This works at the brighter apparent magnitudes,
where the light from the quasar images dominates that from the
galaxy.

Lucey et al (2017) argue that at fainter apparent magnitudes, the
light from the galaxy will dominate that from the quasar.  They report
the discovery of two quadruply lensed quasars that, at first, were
thought to be galaxies.  What singled these objects out was their
differential deblending in the 2MASS and PanSTARRS catalogs.  This
produced astrometric offsets that called for further scrutiny.  Lemon
et al (2017) use differential deblending in the SDSS and GAIA
catalogs, but they start with known quasars.  They might equally well
have started with galaxies.

\section{A Challenge: The Size of Quasar Continuum Emitting Regions}

While we may not be making the same ``completely unproven''
assumptions as other investigators measuring stellar masses,
we have our own set of assumptions.  In particular, we assume
that the continuum emitting region producing the flux ratio anomalies
is sufficiently small -- much smaller than the
Einstein radii of the micro-lenses -- that it can be treated as a point source.

In their original paper, Schechter and Wambsganss (2004) analyzed
optical fluxes and obtained inconsistent results assuming pointlike
emitting regions.  They were able to reconcile those discrepancies
by adopting a toy model in which 50\% of the flux was pointlike and
50\% of the light was very extended and not subject to micro-lensing.

Subsequent work by Pooley et al (2007), Morgan et al (2010) and
Blackburne et al (2011) 
showed that the continuum emitting regions of bright lensed quasars
were factors of 3 - 30 larger than predicted by the venerated Shakura
and Sunyaev (1973) model.

Schechter et al (2014) used X-ray flux ratios in their estimate of the
factor by which Salpteter mass surface densities needed to be
multiplied to allay concerns about the size of the continuum
emitting region.  Jim{\'e}nez-Vicente et al (2015) took a different tack,
carrying out a joint analysis of stellar mass fraction and emitting
region size.  The two approaches yield consistent results, albeit with
large uncertainties.

The size of the X-ray sample will continue to grow as long as the
Chandra X-ray Observatory continues to operate.  Unfortunatly none of
the currently planned X-ray missions will be able to make such
measurements as they lack Chandra's resolution.

As discussed above, newly discovered quadruply lensed quasars are
likely to be less luminous than those first discovered.  It is
reasonable to expect their continuum emitting regions to be
correspondingly smaller, mitigating the effect of their partial
resolution.

\section{Limits on MaCHOs, Including Primordial Black Holes}

In our calculations, we implicitly assume that the dark halo component
of a lens is smoothly distributed.  This translates to halo particles
of at most planetary mass, depending upon the poorly known sizes of
quasar X-ray emitting regions.
\vskip3.0truein
\includegraphics{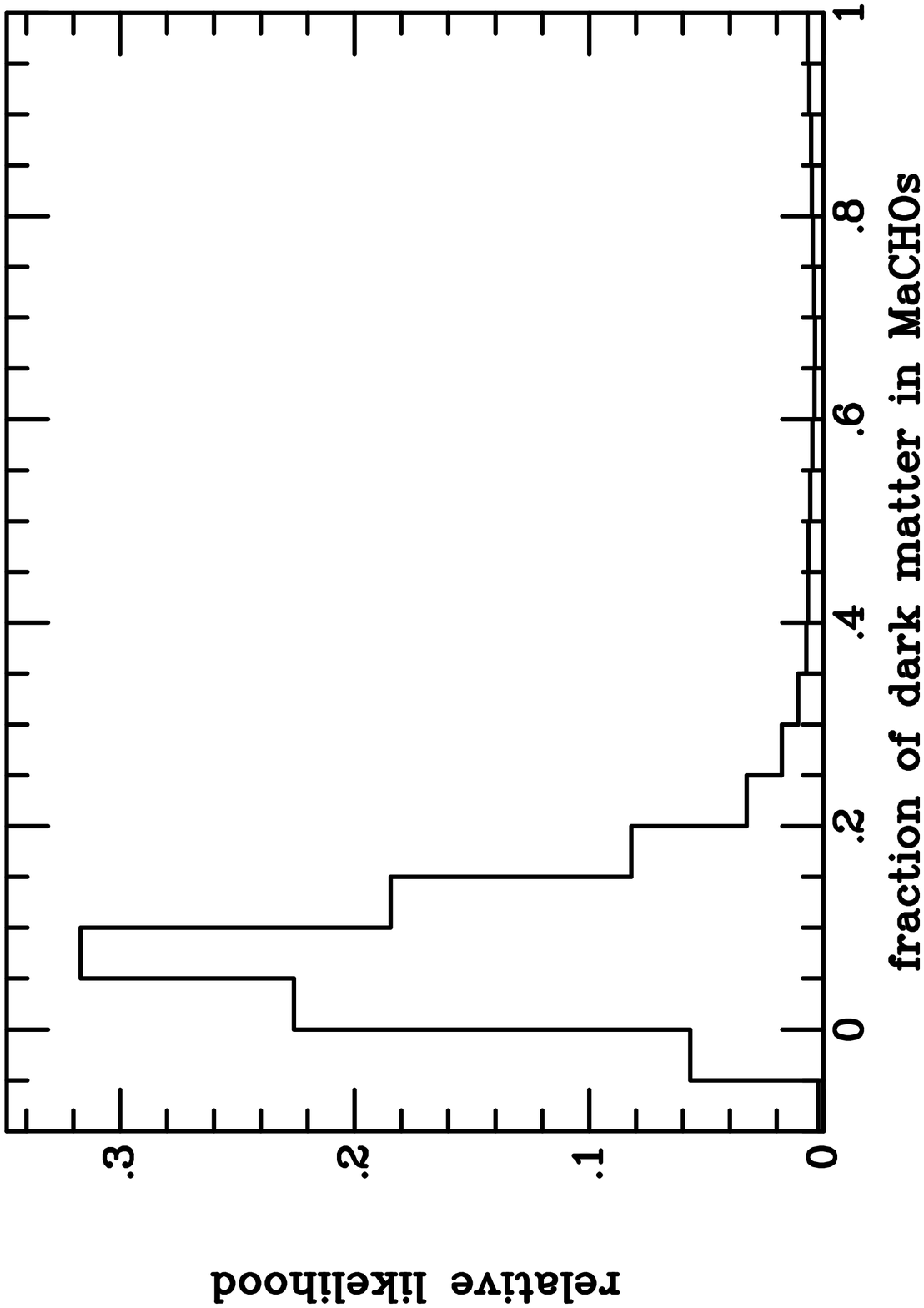}
{\narrower\baselineskip=11pt\noindent
  Figure 3.  Likelihoods for a range of fractional contributions
  of MaCHOs to the dark matter surface density in ten lensed
  quasars.  Note the finite likelihood for a negative fraction,
  which would result if a Salpeter IMF overestimates the surface
  mass density.
  \par}

We can invert our assumptions, and take the stellar surface mass
density to be known (adopting in our case a Salpeter IMF) and instead
let the factor ${\cal F}$ represent the fraction of the dark halo in
Massive Compact Halo Objects (MaCHOs).  The goal is exactly the same
as that of the MaCHO Project (Alcock et al 2000), but we use the
static micro-lensing of quasars rather than the time-variable
micro-lensing of stars.  A significant advantage of the present
technique is that there is no upper limit to masses of the compact
objects.

Mediavilla et al (2009) used static micro-lensing to place explicit
limits on the fraction of halo in the form of primordial black holes,
a subject reviewed by Carr et al (2016).  Their argument was refined
by Mediavilla et al (2017).  They use optical rather than X-ray flux
ratios and the overlap between the two samples is only 50\%, so one
might think it worth the investment of time to re-analyze the Schechter et al
(2014) sample.

The investment was very small.  Exactly one line of code needed to be changed.
Results from that effort are shown if Figure 3.  The most likely
fraction of the dark halo in MaCHOs is something less than 10\%,
confirming the results of Mediavilla et al (2009).  Carl Sagan
famously said ``Extraordinary claims require extraordinary
evidence.''  We suspect Sagan would have preferred to explain
the small excess granularity in lensing galaxies as the product
of a somewhat higher stellar surface mass density.

\acknowledgements The author thanks Jeremy and Joan Mould for many
years of friendship.  Though Jeremy and I collaborated on only one
paper, it was a good one, among the most highly cited publications for
both of us.  By virtue of proximity (we overlapped both in Tucson
and in Pasadena) and affinity we came to know each other well.

Jeremy is unusal among astronomers in that he has always insisted on
thinking things through for himself.  One might have thought this was
among the first requirements for a scientist.  With Jeremy it is actually
the case.  In thinking about our past interactions I can hear his
exaggerated ``hmmmmmmmm'' in response to some new idea or result.  I
can also hear him saying ``It seems to me ...'' followed by a careful
argument.  While I won't get to hear his ``hmmmmmmmm'' when he reads
this contribution, I do look forward to an email beginning ``It seems
to me ...''.



\end{document}